\documentclass[conference]{IEEEtran}
\IEEEoverridecommandlockouts
\usepackage{cite}
\usepackage{amsmath,amssymb,amsfonts}
\usepackage{algorithmic}
\usepackage{textcomp}
\usepackage{xcolor}
\usepackage[pdftex]{graphicx}
\usepackage{epstopdf}

\usepackage{amsmath,amssymb,enumerate} 
\usepackage{bm} 
\usepackage{indentfirst} 
\usepackage{empheq} 
\usepackage{here} 

\newcommand{\ii}{\mathrm{i}}
\newcommand{\nn}{\nonumber}
\newcommand{\Relambda}{{\rm Re}[\lambda_\mu]}
\newcommand{\Imlambda}{{\rm Im}[\lambda_\mu]}
\newcommand{\dmax}{d_{\mathrm{max}}}

\def\BibTeX{{\rm B\kern-.05em{\sc i\kern-.025em b}\kern-.08em
    T\kern-.1667em\lower.7ex\hbox{E}\kern-.125emX}}
\begin{document}

\title{Technology to Counter Online Flaming Based on the Frequency-Dependent Damping Coefficient in the Oscillation Model}

\author{\IEEEauthorblockN{Shinichi Kikuchi}
    \IEEEauthorblockA{
        \textit{Tokyo Metropolitan University} \\
Tokyo 191--0065, Japan \\
kikuchi-shinichi@ed.tmu.ac.jp}
\and
\IEEEauthorblockN{Chisa Takano}
    \IEEEauthorblockA{
        \textit{Hiroshima City University} \\
 Hiroshima, 731--3194 Japan \\
takano@hiroshinma-cu.ac.jp}
\and
\IEEEauthorblockN{Masaki Aida}
    \IEEEauthorblockA{
        \textit{Tokyo Metropolitan University} \\
Tokyo 191--0065, Japan \\
aida@tmu.ac.jp}
}

\maketitle

\begin{abstract}
Online social networks, which are remarkably active, often experience explosive user dynamics such as online flaming, which can significantly impact the real world.
However, countermeasures based on social analyses of the individuals causing flaming are too slow to be effective because of the rapidity with which the influence of online user dynamics propagates.
A countermeasure technology for the flaming phenomena based on the oscillation model, which describes online user dynamics, has been proposed; it is an immediate solution as it does not depend on social analyses of individuals.
Conventional countermeasures based on the oscillation model assume that the damping coefficient is a constant regardless of the eigenfrequency.
This assumption is, however, problematic as the damping coefficients are, in general, inherently frequency-dependent; the theory underlying the dependence is being elucidated.
This paper discusses a design method that uses the damping coefficient to prevent flaming under general conditions considering the frequency-dependence of the damping coefficient and proposes a countermeasure technology for the flaming phenomena.
\end{abstract}

\begin{IEEEkeywords}
online flaming, user dynamics, oscillation model, damping coefficient
\end{IEEEkeywords}

\section{Introduction}
In recent years, with the spread of social networking sites such as Twitter and Facebook, users' activities in online social networks have come to be closely connected to social activities in the real world, not only in online communities.
As a result, the effects of explosive online user dynamics, including the flaming phenomena, are becoming more serious, and countermeasures are needed~\cite{Online_Flaming_com,Online_Flaming_FV}.

Although it is desirable to respond immediately with direct countermeasures to eliminate the factors that cause the flaming phenomena, analyzing each event in detail, one by one, will be too slow to prevent the damage from spreading~\cite{Online_Flaming_youtube,Negative_info_rapid}.
Thus we need an engineering framework for flaming countermeasures that does not depend on the details of each individual event.
One such framework has been proposed~\cite{Network_Dynamics_Introduction}. 
This is based on the oscillation model on networks~\cite{OM_aida_2017,OM_aida_2016} which is used to describe online user dynamics.

Conventional countermeasures for the flaming phenomena have been discussed under the assumption that the damping coefficient \cite{Network_Dynamics_Introduction} is a constant, and independent of the eigenfrequency. 
However, it is known that regardless of the phenomenon, the damping coefficient generally depends on the eigenfrequency.
In fact, the theoretical characteristics of the frequency dependence of the damping coefficient have recently been clarified~\cite{OM_dumping_factor_depends_eigenvalue}. 
Based on these insights, we can consider countermeasures for the flaming phenomena based on the oscillation model, even in general cases where the damping coefficient does depend on the eigenfrequency.

In this paper, we introduce a design methodology that allows the damping coefficient to be used to counter the flaming phenomena even when the damping coefficient depends on the eigenfrequency; we use it to propose a countermeasure technology for the flaming phenomena.

\section{Oscillation Model for Describing Online User Dynamics}
Let the Laplacian matrix of the online social network (OSN) with $n$ nodes be $\bm{\mathcal{L}}$, which is an $n \times n$ square matrix, and the weight of the link from node $i$ to node$j$ $(i \rightarrow j)$ be $w_{ij}$.
In addition, let the eigenvalues of $\bm{\mathcal{L}}$ be $\lambda_\mu$ $(\mu=0,\,1,\,\dots,\,n-1)$ and the eigenvectors associated with $\lambda_\mu$ be $\bm{v}_\mu$.
We assume the eigenvalues are not duplicated.

The eigenvalues of $\bm{\mathcal{L}}$ are generally complex numbers, whose range of existence is given by the largest Gershgorin disk\cite{Gershgorin_theorem} of $\bm{\mathcal{L}}$ as 
\begin{align}
    D_G^{\mathrm{max}}(\bm{\mathcal L}) = \left\{ z \in \mathbb{C} : | z - \dmax | \leq \dmax \right\},
    \label{Gershgorin_disk}
\end{align}
where $\dmax$ is the maximum weighted out-degree of the network.
It is known that all the eigenvalues of $\bm{\mathcal{L}}$ lie within its largest Gershgorin disk~\cite{Network_Dynamics_Introduction}. 

The oscillation model~\cite{Network_Dynamics_Introduction,OM_aida_2017} is a minimal model for describing user dynamics in OSNs.
Let $x_i(t)$ be the state of node (user) $i$ at time $t$. 
Since the influence of interaction between users must propagate through any OSN at a finite speed, its description by the wave equation should be possible, which is the equation for describing the propagation of something at finite speed. 
For the state vector $\bm{x}(t):={}^t\!(x_1(t),\,\dots,\,x_n(t))$, the wave equation in the OSN is written as
\begin{align}
    \frac{{\rm d}^2 }{{\rm d} t^2}\bm x(t) + \bm {\Gamma} \frac{{\rm d}}{{\rm d} t} \bm x(t) = -\bm {\mathcal L} \, \bm x(t),
    \label{Damp_Oscil_Eq}
\end{align}
where $\bm{\Gamma}$ is the matrix expressing the strength of the damping.

Substituting the expansion of $\bm{x}(t)$ by $\bm{v}_\mu$ as 
\begin{align}
    \bm{x}(t) = \sum_{\mu=0}^{n-1} a_\mu(t) \, \bm{v}_\mu ,
    \label{Decomp_Oscil_Mode}
\end{align}
into the wave equation (\ref{Oscil_Mode_Eq}), yields the equation of motion for each oscillation mode $a_\mu(t)$ $(\mu=0,\,1,\,\dots,\,n-1)$ as
\begin{align}
    \frac{{\rm d}^2 }{{\rm d} t^2} a_\mu(t) + \gamma(\omega_\mu) \, \frac{{\rm d}}{{\rm d} t} a_\mu(t) = -\lambda_\mu \, a_\mu(t),
    \label{Oscil_Mode_Eq}
\end{align}
where $\gamma(\omega_\mu)$ is the damping coefficient; it depends on $\omega_\mu=\sqrt{\lambda_\mu}$ and is expressed as 
\[
\gamma(\omega_\mu) := \gamma_0 + \gamma_1\,\lambda_\mu
\]
with the constant $\gamma_0$ and $\gamma_1$~\cite{OM_dumping_factor_depends_eigenvalue}.
Note that $\mathrm{Re}[\gamma(\omega_\mu)]=\gamma_0 + \gamma_1 \, \Relambda \geq 0$.

The solution of (\ref{Oscil_Mode_Eq}) is written as
\begin{align}
    a_\mu(t) &= c_\mu^+ \exp\!\left[ -\frac{\gamma(\omega_\mu)}{2}\,t + \ii \,\sqrt{r_\mu} \, \exp{\!\left(\ii\,\frac{\theta_\mu}{2}\right)}\,t \right] \nonumber \\
                    &\ \ \ \ +  c_\mu^- \exp{\!\left[ -\frac{\gamma(\omega_\mu)}{2}\,t - \ii \,\sqrt{r_\mu} \, \exp{\!\left(\ii\,\frac{\theta_\mu}{2}\right)}\,t \right] },
\end{align}
where $c_\mu^+$ and $c_\mu^-$ are constants that depend on $\mu$, 
and $r_\mu$ and $\theta_\mu (-\pi < \theta_\mu \leq \pi)$ are, respectively, the absolute value and the argument of the following complex number: 
\begin{align}
    r_\mu \,\exp(\ii \,\theta_\mu) := \lambda_\mu - \left( \frac{\gamma(\omega_\mu)}{2} \right)^2 = - \left( \alpha + \frac{\gamma(\omega_\mu)}{2} \right)^2.
    \label{polor_rep}
\end{align}

In the oscillation model, the oscillation energy can be considered as the strength of the activity of user dynamics \cite{OM_node_centrality,OM_node_centrality2}.
Also, the situation in which oscillation energy diverges over time is considered to describe explosive user dynamics, which include the flaming phenomena.
Therefore, in order to prevent explosive user dynamics, it is necessary to consider the conditions under which the oscillation energy does not diverge.
By deriving the strength of the damping that satisfies this condition, we can obtain a framework in which the strength of damping can be adjusted to prevents the flaming phenomena.

The conventional solutions to the flaming phenomena assume that the damping coefficient is a constant and independent of eigenfrequency.
This corresponds to the special case of $\gamma_1=0$ in (\ref{Oscil_Mode_Eq}).
Following \cite{Network_Dynamics_Introduction}, the condition under which the oscillation energy does not diverge is given as
\begin{align}
    {}^\forall \mu, \quad \frac{\gamma_0}{2\,\sqrt{r_\mu}} \geq \left|  \sin{\frac{\theta_\mu}{2}}  \right|,
    \label{Cond_non_diverge_sin}
\end{align}
and the value of the damping coefficient required to satisfy this condition is given by
\begin{align}
    \gamma_0 \geq \sqrt{2\,\dmax}.
\end{align}

In the next section, in order to consider countermeasures for the flaming phenomena, we discuss the conditions under which the oscillation energy does not diverge in the case of $\gamma_1\not= 0$. 

\section{Model of Explosive User Dynamics Considering Frequency-Dependent Damping Coefficient} 
Since the oscillation energy is proportional to the square of the absolute value of $a_\mu(t)$, we derive the condition under which $a_\mu(t)$ does not diverge and then the condition under which flaming does not occur.

By decomposing the damping coefficient $\gamma(\omega_\mu)$ into real and imaginary parts as in
\begin{align}
\gamma(\omega_\mu) = \left(\gamma_0 + \gamma_1 \, \Relambda\right) + \ii \, \gamma_1 \, \Imlambda .
\end{align}
$a_\mu(t)$ is written as
\begin{align}
    a_\mu(t) =& \  c_\mu^+ \exp{\!\left[ -\frac{\gamma_0 + \gamma_1\Relambda}{2}t - \sqrt{r_\mu} \sin{\!\left(\frac{\theta_\mu}{2}\right) t}\right] } \nn \\
                    & \ \ \times c_\mu^+ \exp{\!\left[ -\ii\frac{\gamma_1\Imlambda}{2}t + \ii \sqrt{r_\mu} \cos{\!\left(\frac{\theta_\mu}{2}\right) t} \right] } \nn \\
                    &+ \ c_\mu^- \exp{\!\left[ -\frac{\gamma_0 + \gamma_1\Relambda}{2}t + \sqrt{r_\mu} \sin{\!\left(\frac{\theta_\mu}{2}\right) t} \right] } \nn \\
                    & \ \ \times c_\mu^- \exp{\!\left[ -\ii\frac{\gamma_1\Imlambda}{2}t - \ii \sqrt{r_\mu} \cos{\!\left(\frac{\theta_\mu}{2}\right) t} \right] }.
\label{eq:a_u(t)}
\end{align}
To determine if $a_\mu(t)$ diverges or not, we need to check whether the real components of the exponent of the exponential function in (\ref{eq:a_u(t)}) are positive or negative, and if $a_\mu(t)$ diverges, the following condition is satisfied: 
\begin{align*}
    \!\!\!\!
    \left( \frac{\mathrm{Re}[\gamma(\omega_\mu)]}{2\,\sqrt{r_\mu}} +  \sin{\!\left(\frac{\theta_\mu}{2}\right)} \right)
    \left( \frac{\mathrm{Re}[\gamma(\omega_\mu)]}{2\,\sqrt{r_\mu}} -  \sin{\!\left(\frac{\theta_\mu}{2}\right)} \right) < 0. 
\end{align*}
This means the one of real components of the exponent is positive and the other is negative. 

Consequently, the condition under which the oscillation energy diverges is given by
\begin{align}
    {}^\exists \mu, \quad \frac{\gamma_0 + \gamma_1\Relambda}{2\,\sqrt{r_\mu}} < \left|  \sin{\frac{\theta_\mu}{2}}  \right|,
    \label{振動エネルギーが発散する条件}
\end{align}
and the condition that the oscillation energy does not diverge is obtained as
\begin{align}
    {}^\forall \mu, \quad \frac{\gamma_0 + \gamma_1\Relambda}{2\,\sqrt{r_\mu}} \geq \left| \sin{\frac{\theta_\mu}{2}} \right| .
    \label{cond_non_div}
\end{align}
Since the conventional condition (\ref{Cond_non_diverge_sin}) that the oscillation energy does not diverge corresponds to the case of $\gamma_1=0$ as per condition (\ref{cond_non_div}), the condition (\ref{cond_non_div}) for the frequency-dependent damping coefficient is a generalization of the conventional result.

\section{Countermeasure for Flaming Phenomena Given the Frequency-Dependent Damping Coefficient}
Based on condition (\ref{cond_non_div}), i.e., the oscillation energy does not diverge, we consider a design method for the damping coefficient to satisfy (\ref{cond_non_div}), and consider a countermeasure technology for the flaming phenomena by adjusting the value of the damping coefficient.

\subsection{Adjusting the Damping Coefficient}
Among parameters $\gamma_0$ and $\gamma_1$, which determine the strength of damping, $\gamma_1$ is the eigenfrequency dependent term. 
This means that the value of $\gamma_1$ is a parameter predetermined by the structure of the network. 
Therefore, $\gamma_0$ is the only parameter that can be manipulated independently of the network structure.
In this framework, even if various values are given as $\gamma_1$, we can counter the flaming phenomena by adjusting the value of $\gamma_0$ . 
Here, the actual action to adjust the value of $\gamma_0$ includes disseminating other information to attract users' attention.
In the following, we consider the value of $\gamma_0$ necessary to prevent the triggering of explosive user dynamics, and we use its value in a flaming countermeasure.

The range of eigenvalues of $\bm{\mathcal{L}}$ is the interior of the largest Gershgorin disk (including its boundaries) of radius $\dmax$ with center $(\dmax,0)$.
From condition (\ref{cond_non_div}), the oscillation energy does not diverge, we can consider the range satisfying condition (\ref{cond_non_div}) on the complex plane. 
Then, if the largest Gershgorin disk of $\bm{\mathcal{L}}$ lies completely within the range on the complex plane, the oscillation energy never diverges regardless of the network structure.

In order to clarify the region on the complex plane in which condition (\ref{cond_non_div}) ensures that the oscillation energy does not diverge, the inequality of condition (\ref{cond_non_div}) is transformed as follows: 
\begin{align}
    \Imlambda^2  \leq \frac{Z}{ 4 - 4\gamma_0 \, \gamma_1 - 4 \, \gamma_1^2 \, \Relambda },
    \label{cond_non_div_complex}
\end{align}
where
\begin{align*}
    Z &= \gamma_0^4 + 4\gamma_0^3 \, \gamma_1 \, \Relambda + 6 \, \gamma_0^2 \, \gamma_1^2 \, \Relambda^2 \\
    &\ \ \ + 4 \, \gamma_0^2 \, X + 4 \, \gamma_0 \, \gamma_1^3 \, \Relambda^3 + 8 \, \gamma_0 \, \gamma_1 \, \Relambda X \\
    &\ \ \ + \gamma_1^4 \, \Relambda^4 + 4 \, \gamma_1^2 \, \Relambda^2 \, X,
\end{align*}
and $X$ is the real part of (\ref{polor_rep}).

The range of eigenvalues of $\bm{\mathcal{L}}$, determined by the Gershgorin theorem, are written as 
\begin{align}
    \Imlambda^2 \leq \dmax^2 - (\Relambda - \dmax)^2 .
    \label{Gershgorin_2}
\end{align}
We compare (\ref{Gershgorin_2}) and the condition (\ref{cond_non_div_complex}) that the oscillation energy does not diverge.
Because both of them are axially symmetric on by the real axis, we consider the upper-half plane. 
The condition that the largest Gershgorin disk is completely included the range of (\ref{cond_non_div_complex}) is expressed as 
\begin{align}
    \!\!
    \dmax^2 - (\Relambda - \dmax)^2 \leq \frac{Z}{4\,(1 - \gamma_0\,\gamma_1 - \gamma_1^2\,\Relambda) }, 
    \notag
\end{align}
which can be transformed to
\begin{align}
    &\left( \gamma_0 \, \gamma_1 + 1 + 2 \, \gamma_1^2 \, \dmax \right)\Relambda \nn \\
    &\quad \geq -(\gamma_0^2 + 2 \, \gamma_0 \, \gamma_1\,\dmax - 2 \, \dmax ), 
    \label{Gershgorin_enegy_non-div2}
\end{align}
by considering $\Relambda \ge 0$.

We consider the conditions for satisfying inequality (\ref{Gershgorin_enegy_non-div2}) in the following three cases.
\begin{itemize}
\item if $\gamma_0\gamma_1 + 1 + 2\gamma_1^2\,\dmax = 0$,
\begin{align}
    \gamma_0^2 + 2 \, \gamma_0 \, \gamma_1\,\dmax - 2 \, \dmax \ge 0.
\end{align}
The range of $\gamma_0$ that satisfies the above is as follows from $\gamma_0 \ge 0$
\begin{align}
    \gamma_0 \ge \sqrt{\gamma_1^2\,\dmax^2 + 2\,\dmax} - \gamma_1\,\dmax .
\end{align}
\item if $\gamma_0\gamma_1 + 1 + 2\gamma_1^2\,\dmax > 0$, 
\begin{align}
 \frac{ \gamma_0^2 + 2\,\gamma_0\,\gamma_1\,\dmax - 2\,\dmax }{\gamma_0\,\gamma_1 + 1 + 2\,\gamma_1^2\,\dmax} &\geq - \Relambda.
\label{cond_non_div_complex_with_gamma_0_gamma_1_d_max}
\end{align}
In order for this inequality to hold, the numerator needs to be non-negative, so we obtain  
\begin{align}
\gamma_0^2 + 2\,\gamma_0\,\gamma_1\,\dmax - 2\,\dmax \geq 0. 
\label{case2}
\end{align}
Considering $\gamma_0 \geq 0$, the condition of $\gamma_0$ to ensure the non-divergence of oscillation energy for all eigenvalues is written as
\begin{align}
\gamma_0 \geq \sqrt{\gamma_1^2 \, \dmax^2 + 2 \, \dmax} - \gamma_1 \, \dmax .
\label{condition of gamma_0 : gamma_1>0}
\end{align}
\item if $\gamma_0\,\gamma_1 + 1  + 2\,\gamma_1^2\,\dmax < 0$,
\begin{align}
    \frac{ \gamma_0^2 + 2\,\gamma_0\,\gamma_1\,\dmax - 2\,\dmax }{\gamma_0\,\gamma_1 + 1 + 2\,\gamma_1^2\,\dmax} &\leq - \Relambda.
    \label{case3}
\end{align}
Inequality (\ref{case3}) is transformed to
\begin{align}
 (\gamma_0 + 2 \, \gamma_1 \, \dmax)^2 &\geq 0
\end{align}
Therefore, inequality (\ref{case3}) always holds in this case.

\end{itemize} 

To summarize the above results, the $\gamma_0$ condition that ensures the oscillation energy does not diverge for all eigenvalues is obtained as
\begin{align}
    \gamma_0 \geq \sqrt{\gamma_1^2 \, \dmax^2 + 2 \, \dmax} - \gamma_1 \, \dmax .
    \label{condition of gamma_0}
\end{align}

Therefore, given the maximum weighted out-degree of the network, $\dmax$, and parameter $\gamma_1$ of the damping coefficient, adjusting the value of $\gamma_0$ to satisfy (\ref{condition of gamma_0}) will counter the flaming phenomena.

\subsection{Case Studies}
Using an example network with $\dmax=100$, this section considers three cases of different values of $\gamma_1$, the frequency dependence of the damping coefficient: $\gamma_1=0$, $\gamma_1>0$ or $\gamma_1<0$.
In all cases, we confirm that the region of the condition that the oscillation energy does not diverge includes the largest Gershgorin disk of $\bm{\mathcal{L}}$ by satisfying the condition  (\ref{condition of gamma_0}) of $\gamma_0$.
\begin{figure*}[t]
    \centering
    \includegraphics[width=0.6\linewidth]{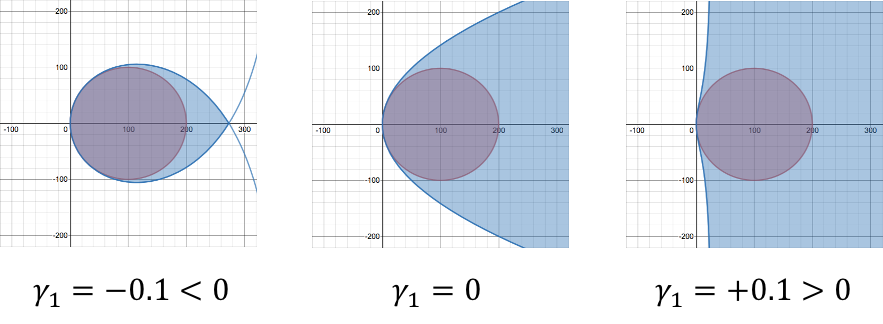}
    \caption{Examples of complete flaming prevention}
    \label{fig:prevent}
\end{figure*}
\begin{figure*}[t]
    \centering
    \includegraphics[width=0.6\linewidth]{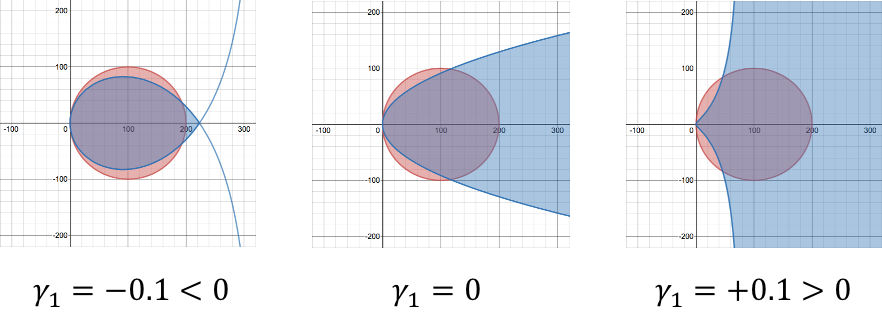}
    \caption{Examples of incomplete flaming prevention}
    \label{fig:flaming}
\end{figure*}

First, we confirm the case of complete flaming prevention with $\gamma_0 = \sqrt{\gamma_1^2 \, \dmax^ 2 + 2 \, \dmax} - \gamma_1\,\dmax$, where $\gamma_0$ is the minimum value that satisfies condition (\ref{condition of gamma_0}).
Figure \ref{fig:prevent} shows the regions in which the oscillation energy does not diverge as given by condition (\ref{cond_non_div}), for the cases of $\gamma_1 = 0.1$, $\gamma_1 = 0$ and $\gamma_1 = -0.1$.
These regions are depicted in blue. 
In addition, the largest Gershgorin disk of $\bm{\mathcal{L}}$ is depicted in red.
In all figures, it can be seen that the regions wherein the oscillation energy does not diverge completely include the largest Gershgorin disk, so that no divergence of oscillation energy occurs regardless of the details of the network structure.

Next, we show the case of incomplete flaming prevention by using $\gamma_0 = \sqrt{\gamma_1^2 \, \dmax^ 2 + 2 \, \dmax} - \gamma_1\,\dmax - 5$, in which $\gamma_0$ is less than the minimum value that satisfies condition (\ref{condition of gamma_0}).
Figure \ref{fig:flaming} shows the regions wherein the oscillation energy does not diverge as indicated by condition (\ref{cond_non_div}), for the cases of $\gamma_1 = 0.1$, $\gamma_1 = 0$, and $\gamma_1 = -0.1$.
In these cases, the regions cannot completely enclose the Gershgorin disk.
If even just one eigenvalue appears outside of the region, the oscillation energy diverges and the flaming phenomenon occurs.
Therefore, depending on the position of the eigenvalues of $\bm{\mathcal{L}}$, flaming prevention is not assured.

\section{Conclusion}
In this paper, we proposed a countermeasure technology for the flaming phenomena based on the oscillation model with the frequency-dependent damping coefficient.
The design method that yields the damping coefficients using condition (\ref{condition of gamma_0}) is a generalized version of the conventional countermeasure technology for the flaming phenomena.
Regardless of the value of parameter $\gamma_1$, which is the strength of the frequency dependence of damping coefficient, we can prevent explosive user dynamics by setting parameter $\gamma_0$ to be an appropriate value. 

Furthermore, the required value of $\gamma_0$ can be determined from just $\dmax$, which is the maximum weighted out-degree of the network, and $\gamma_1$, which is the strength of the frequency-dependence of the damping coefficient.
One of the methods for increasing the value of $\gamma_0$ in the actual OSNs is that to disseminate other information to attract users' attention is mentioned.

\section*{Acknowledgement}
This research was supported by Grant-in-Aid for Scientific Research 19H04096 and 20H04179 from the Japan Society for the Promotion of Science (JSPS).


\end{document}